# Elastic properties of self-folded two-dimensional nanomaterials: a theoretical model validated by molecular dynamics simulations

Anran Wei [a], Fenglin Guo [a][*]

[a] School of Naval Architecture, Ocean and Civil Engineering (State Key Laboratory of Ocean Engineering), Shanghai Jiao Tong University, Shanghai 200240, China

* Corresponding author, E-mail: *flguo@sjtu.edu.cn*

**Abstract**

The trade-off between strength and ductility has plagued the design of macroscopic assemblies of two-dimensional materials for a long time. In order to break the strength-ductility paradox, the design of self-folded two-dimensional nanomaterial (SF-2DNM) has been recently proposed with the inspiration from the folded nanostructures of natural silks. Such folding strategy is revealed to greatly enhance the ductility of overall assembly without much sacrifice of the excellent tensile strength of two-dimensional materials. However, the dependences of the elastic properties of SF-2DNMs on the material properties of building blocks and the geometries of folded structures have not been specifically clarified in previous studies. In this paper, we thus develop a theoretical model to describe the elastic properties of SF-2DNMs based on the shear-lag analysis. The load transfer behaviors and failure modes of SF-2DNMs are demonstrated with this model. The Young's modulus and tensile strength of SF-2DNMs are also predicted, further validated by molecular dynamics simulations. This model brings insights into the elastic deformation of SF-2DNMs under external tension. The structure-property relationship revealed by this model would provide useful guidelines for the rational design of SF-2DNMs in engineering applications.

## 1. Introduction

In recent decades, researches on the two-dimensional nanomaterials (2DNMs) have risen rapidly due to their exceptional physical properties [1-3]. The 2DNMs are usually gifted with extraordinary mechanical properties comparing with conventional materials. It is reported that the graphene as a well-known example of 2DNMs shows amazing Young's modulus of ~1 TPa and tensile strength of ~120 GPa [4]. Hereafter, huge efforts have been devoted to constructing macroscopic materials with excellent mechanical performances by the assembling of 2DNMs for various engineering applications, such as the 2DNM-architected films [5, 6], papers [7, 8], fibers [9, 10], aerogels [11], etc. However, the mechanical properties of these macroscopic assemblies of 2DNMs are commonly far poorer than their building blocks in nanoscale. One of the long-troubled problems is the trade-off between the strength and ductility. High ductility is pursued for many 2DNM-based materials that require good energy dissipation [12], flexibility [13] or shape programmability [14] in their applications. At the expense of high ductility, the substantial sacrifice of strength is usually unavoidable. For example, Xiao et al. [15] fabricated a rubber-like graphene film with a significantly large fracture strain of 23%, but its tensile strength only reaches ~30 MPa, which is much poorer than the ideal value of graphene nano-flake.

In order to break such strength-ductility paradox, a bio-inspired design of the nanostructures in 2DNM-based materials has been proposed by Jia et al. [16] recently. The evolution over billions of years brings many strong and extensible biomaterials. A typical example is the spider silk, as shown in Fig. 1a. It is revealed that the folded $\beta$- sheet nanocrystals embedded in semi-amorphous protein matrix give both high strength and ductility to the spider silk [17]. Inspired by the self-folded nanostructures in protein, Jia et al. [16] developed a structure design of self-folded 2DNMs (SF-2DNMs). In their design, the layers of 2DNM (taking graphene as an example) are architected to be regularly folded with a uniform fold length in nanoscale and assembled into a macroscopic material, as illustrated in Fig. 1b. They adopted detailed simulations to verify that the design of self-folded nanostructures yields ultrahigh ductility of SF-2DNMs while preserving a large amount of tensile strength of 2DNMs. Moreover, the mass fabrication and precious size control of folded nanostructures have been successfully achieved in 2DNM by experiments [13, 18]. It is suggested that the structure design by this folding strategy is highly promising to fabricate strong-yet-ductile 2DNM-based materials. Therefore, understanding the relationship between the folded nanostructure and overall mechanical properties is urged for the material design and engineering applications. Accompanied with the folding strategy, an analytical expression was also provided by Jia et al. [16] for effective predictions of the ductility of $n$-folded SF-2DNMs. However, the clear and direct descriptions of the structure-dependent elastic properties of SF-2DNMs remain unclear, which was only roughly modeled and discussed in their study.

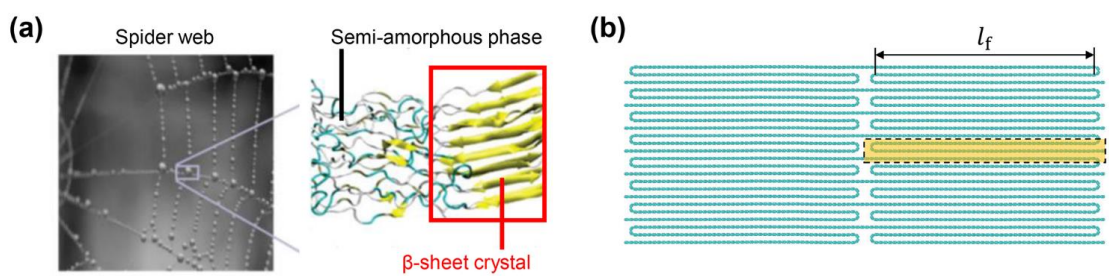

**Fig. 1 The design of self-folded two-dimensional nanomaterials (SF-2DNMs) inspired by the**

**folded nanostructure of natural silks.** (**a**) The photo of natural spider silk with the illustration of folded $\beta$-sheet nanocrystals embedded in semi-amorphous protein matrix (adapted with permission from [17]. Copyright 2010, American Chemical Society). (**b**) The folded structure of SF-2DNMs with a regular pattern and uniform fold length $l_\mathrm{f}$. The representative volume element (RVE) for the structure is indicated by the yellow dashed box.

In this paper, we thus develop a theoretical model to investigate the effects of structural geometries on the overall elastic properties of SF-2DNMs, based on the shear-lag analysis in continuum mechanics. The shear-lag method is used to model the mechanical properties of fiber-reinforced composites [19] and nacre-like biomaterials [20]. Meanwhile, it has been widely adopted for the mechanical modeling of layer-by-layer stacked assemblies of 2DNMs. Liu et al. [21] firstly developed a deformable tension-shear (DTS) model for the layered assemblies of graphene by considering the in-plane deformation in the classic shear-lag model. With the DTS model, Gao et al. [22] studied the optimization of elastic properties by manipulating the interlayer crosslinks between stacked graphene layers. Further, the non-linear DTS model is generated for the elastic-plastic constitutive relationship of interlayer crosslinks [23, 24]. Moreover, Wei et al. [25] improved the DTS model by considering the statistical phenomenon in the material parameters and geometrical sizes. Here, we also apply the shear-lag analysis for SF-2DNMs to derive a model targeted for self-folded nanostructures. Our model demonstrates the load transfer behaviors in SF-2DNMs. The elastic properties of SF-2DNMs, such as Young's modulus and tensile strength, can be well predicted by our model, which is also validated by the molecular dynamics (MD) simulations. Our model can serve as a guideline for the rational design and manipulation of SF-2DNMs.

## 2. Theoretical modeling

By neglecting the diversity and non-uniformity of folded nanostructures, we model the SF-2DNM as a regularly folded structure with identical fold length $l_\text{f}$ and periodic pattern. For simplification, we consider the most basic tri-folded configuration in this model. Then the periodicity and symmetry of the structure allow us to analyze the mechanical properties of overall SF-2DNM by a representative volume element (RVE) shown in Fig. 2.

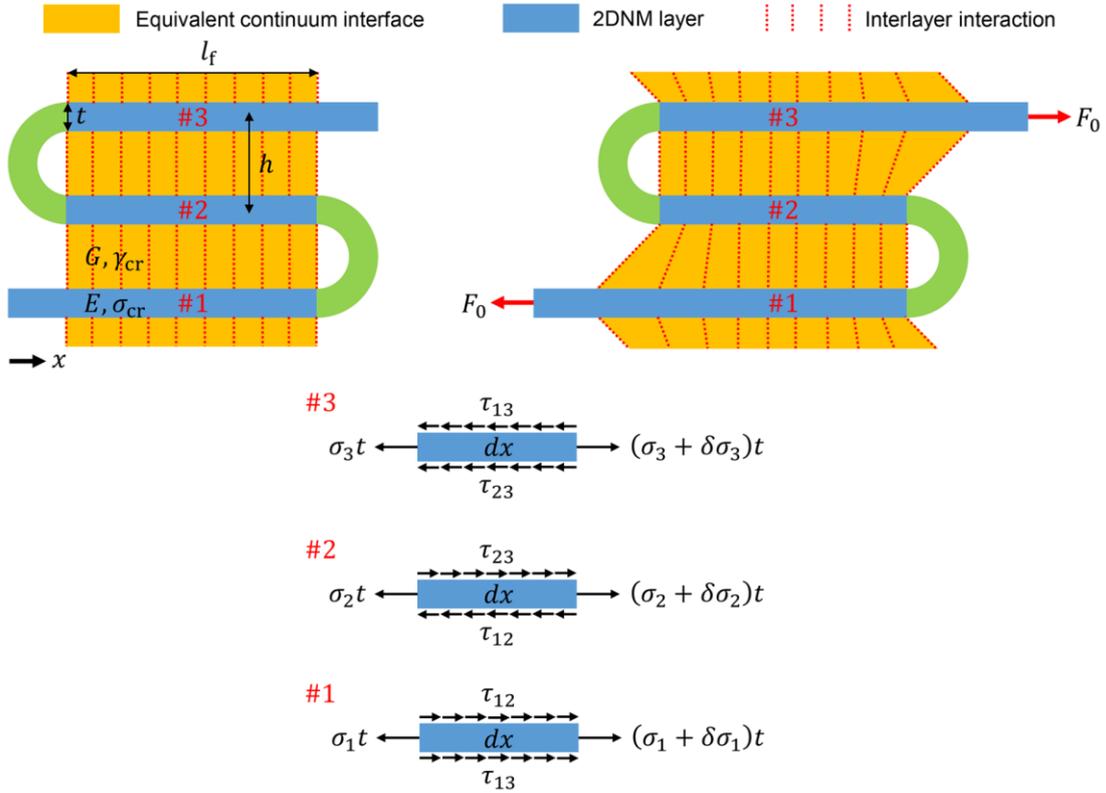

**Fig. 2 Schematic illustrations of the theoretical model for self-folded two-dimensional nanomaterials (SF-2DNMs).** A representative volume element (RVE) with a basic tri-folded configuration is adopted in our model. The interlayer interaction is represented by an equivalent continuum layer. The three folded 2DNM layers are marked as #1, #2 and #3. The material

properties and structural geometries are also labeled as those defined in the model. The analysis of stress equilibrium is conducted for each layer.

In the RVE, the SF-2DNM can be modeled by three parallel layers denoted by #1, #2, #3 with rigid connections applied between the edges of layers. Here, the elastic deformation of the connections between layers is ignored, since these connections are the secondary structures comparing the stacked layers with large aspect ratios. These layers with an interlayer distance $h$ are regarded as a purely elastic material with a stiffness $D = Et$, where $E$ is the Young's modulus and $t$ is the thickness of layer, and a tensile strength $\sigma_{cr}$. In addition to the covalent bonds in the material layers, the interlayer van der Waals interactions also contribute to the mechanical properties of SF-2DNM. Since the interlayer interactions are shearing dominant when a remote tensile load is applied, an artificial continuum layer is introduced between layers #1, #2 and #3 to model the interactions between adjacent layers. This artificial layer is assumed to be elastic and brittle with shear modulus $G$ and critical shear strain $\gamma_{cr}$.

For layers #1, #2 and #3, the equilibrium equations can be derived by taking an infinitesimal length $dx$ for analysis, as illustrated in Fig. 2. Considering the in-plane and interlayer displacements, one can get the governing equations as followings

$$\begin{cases} D\dfrac{\partial^2 u_1}{\partial x^2} = -\tau_{12} - \tau_{13} = \dfrac{G}{h}(2u_1 - u_2 - u_3) \\ D\dfrac{\partial^2 u_2}{\partial x^2} = \tau_{12} - \tau_{23} = \dfrac{G}{h}(2u_2 - u_1 - u_3) \\ D\dfrac{\partial^2 u_3}{\partial x^2} = \tau_{13} + \tau_{23} = \dfrac{G}{h}(2u_3 - u_1 - u_2) \end{cases} \quad (1)$$

where $u_i$ (i = 1, 2, 3) is the in-plane displacement profile in layer #i, and $\tau_{ij}$ (i, j = 1, 2, 3) is the shear stress between layer #i and #j which direction is defined as the schematics plotted in Fig. 2. The tensile loads acting on both ends of the RVE is $F_0$, which should be equal to the tensile forces in the layer #1 at the boundary of $x = 0$ and layer #3 at the boundary of $x = l_f$. That is written into the conditions of

$$D\frac{\partial u_1(0)}{\partial x} = F_0, \qquad D\frac{\partial u_3(l_f)}{\partial x} = F_0 \tag{2}$$

The rigid connections between the edges of layers require the constraints of stress and displacement at the edges of different layers as the forms of

$$\frac{\partial u_1(l_f)}{\partial x} = -\frac{\partial u_2(l_f)}{\partial x}, \qquad \frac{\partial u_2(0)}{\partial x} = -\frac{\partial u_3(0)}{\partial x} \tag{3}$$
$$u_1(l_f) = u_2(l_f), \qquad u_2(0) = u_3(0)$$

Combining Eqs. (1-3), the in-plane displacement in the three layers can be solved as

$$\begin{cases} u_1 = A_1 + \dfrac{F_0 l_0}{3D}\left[\dfrac{x}{l_0} - \dfrac{(n-2)e^{x/l_0} - (n-2n^2)e^{-x/l_0} + n^2 - 1}{n^2 - n + 1}\right] \\ u_2 = A_1 + \dfrac{F_0 l_0}{3D}\left[\dfrac{x}{l_0} - \dfrac{(n+1)e^{x/l_0} - (n^2+n)e^{-x/l_0} + n^2 - 1}{n^2 - n + 1}\right] \\ u_3 = A_1 + \dfrac{F_0 l_0}{3D}\left[\dfrac{x}{l_0} - \dfrac{(1-2n)e^{x/l_0} - (n^2-2n)e^{-x/l_0} + n^2 - 1}{n^2 - n + 1}\right] \end{cases} \tag{4}$$

Here, $A_1$ is a parameter related to the rigid body displacement of the overall system, $l_0 = \sqrt{Dh/3G}$ is the characteristic length that is determined by the intrinsic properties of a certain 2DNM and introduced for the normalization of length in our following discussions, $n = e^{l_f/l_0}$ is a nondimensional parameter containing the normalized fold length $l_f/l_0$. This solution of displacement is then used for the detailed analysis of the overall elastic properties of SF-2DNM in our next section.

### 3. Results and discussions

*3.1 Load transfer behaviors in SF-2DNMs*

In the model of SF-2DNM under remote tension, the external load is transferred not only by the lattice stretch in the folded material layers but also the shearing deformation between adjacent layers. Investigations on the load transfer behaviors can give us an in-depth understanding on the overall elastic properties of SF-2DNM. The intralayer and interlayer loading status can be observed from the tensile force distribution $F(x)$ and the shear strain distribution $\gamma(x)$ in each layer. For an arbitrary layer #i, the tensile force is calculated as $F_i = D\partial u_i/\partial x$, which is expressed as

$$\begin{cases} F_1 = \dfrac{F_0}{3}\left[1 - \dfrac{(n-2)e^{x/l_0} + (n-2n^2)e^{-x/l_0}}{n^2 - n + 1}\right] \\ F_2 = \dfrac{F_0}{3}\left[1 - \dfrac{(n+1)e^{x/l_0} + (n^2+n)e^{-x/l_0}}{n^2 - n + 1}\right] \\ F_3 = \dfrac{F_0}{3}\left[1 - \dfrac{(1-2n)e^{x/l_0} + (n^2-2n)e^{-x/l_0}}{n^2 - n + 1}\right] \end{cases} \quad (5)$$

Meanwhile, the interlayer shear strain can be derived by $\gamma_{ij} = (u_j - u_i)/h$, following the forms of

$$\begin{cases} \gamma_{12} = \dfrac{F_0 l_0}{Dh} \cdot \dfrac{-e^{x/l_0} + n^2 e^{-x/l_0}}{n^2 - n + 1} \\ \gamma_{13} = \dfrac{F_0 l_0}{Dh} \cdot \dfrac{(n-1)e^{x/l_0} + (n^2-n)e^{-x/l_0}}{n^2 - n + 1} \\ \gamma_{23} = \dfrac{F_0 l_0}{Dh} \cdot \dfrac{ne^{x/l_0} - ne^{-x/l_0}}{n^2 - n + 1} \end{cases} \quad (6)$$

where the subscripts i and j are used to represent the shear strain between layers #i and #j. From these analytical results, we notice that SF-2DNMs with varying fold lengths $l_f/l_0$ show different profiles of $F_i(x)$ and $\gamma_{ij}(x)$. We then plot the distributions of in-plane tensile force and interlayer shear strain for the SF-2DNM models with $l_f/l_0 = $ 3, 5, 10 and 20, as shown in Fig. 3 and Fig. 4, respectively. Due to the structural symmetry of RVE, $F_1(x)$ and $F_3(x)$ are symmetric about the central axis $x = l_f/2$, and $F_2(x)$ is self-symmetric about the same axis. Similarly, $\gamma_{12}(x)$ and $\gamma_{23}(x)$ are mirrored as well as $\gamma_{13}(x)$ shows the same self-symmetry. These solutions of force and shear strain distributions follow the above boundary conditions.

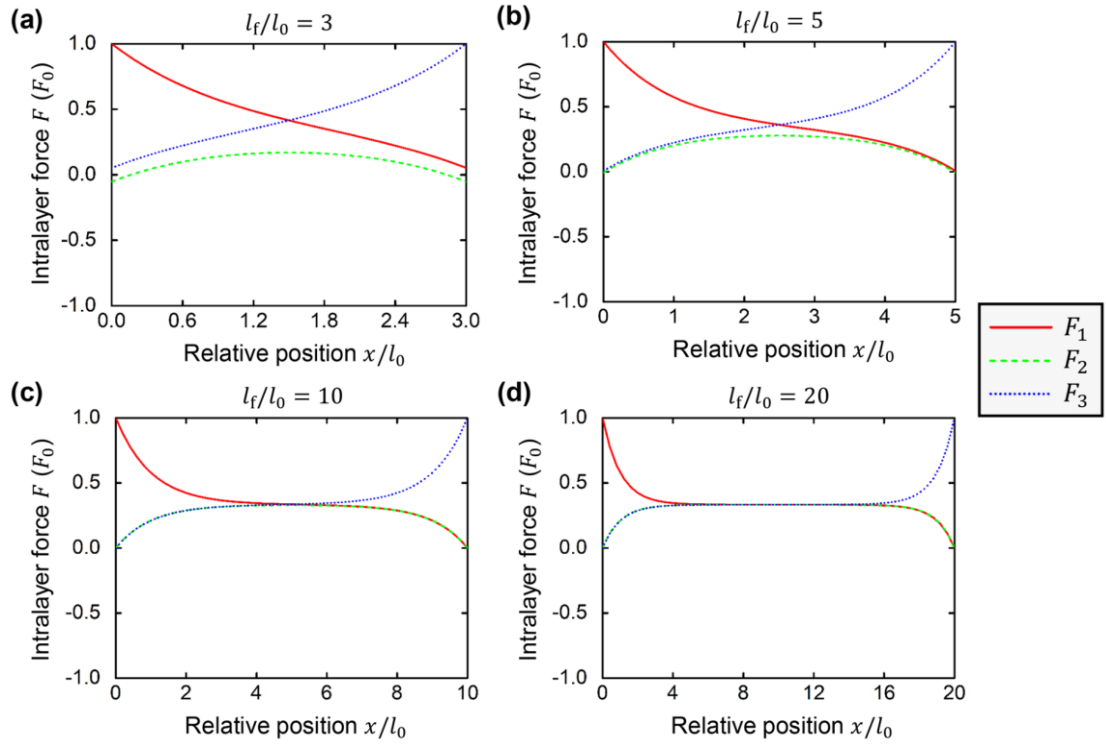

Fig. 3 Distributions of tensile force in self-folded two-dimensional nanomaterials (SF-2DNMs) with normalized fold length $l_f/l_0 = 3, 5, 10$ and $20$.

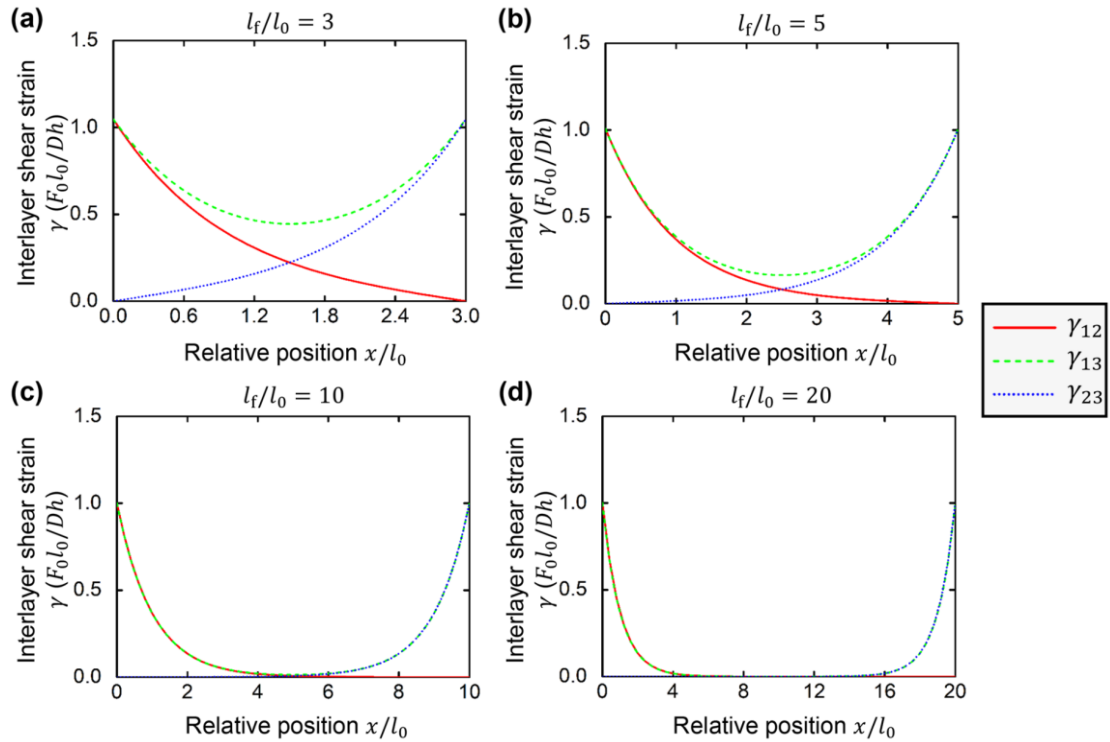

Fig. 4 Distributions of interlayer shear strain in self-folded two-dimensional nanomaterials

(SF-2DNMs) with normalized fold length $l_f/l_0 = 3, 5, 10$ and $20$.

Interestingly, the profiles of $F(x)$ and $\gamma(x)$ greatly changes with the fold length $l_f/l_0$. When $l_f/l_0$ is relatively small, for example $l_f/l_0 = 3$, the tensile forces in layers #1 and #3 are nearly linear along the length direction and quite distinct from that in layer #2, as shown in Fig. 3a. However, when $l_f/l_0$ is relatively large, the linearity of tensile force is only limited within the regions near the edges of layers, and the central regions of the profile of $F(x)$ are flattened, as shown in Fig. 3b-d. Such change indicates that the load transfer between adjacent layers through interlayer shearing is significant weakened with the increasing fold length. That can be also reflected from the shear strain that is gradually vanished with the increasing $l_f/l_0$ in the center regions of layers #1, #2 and #3, as shown in Fig. 4b-d. It can be estimated from these curves that most interlayer load transfer is conducted only within the distance of $\sim 4l_0$ from the edges of layer no matter how long the layers are. Besides, with the increasing $l_f/l_0$, the profile of $F_2(x)$ progressively overlaps with those of $F_1(x)$ and $F_3(x)$ at regions of $x > l_f/2$ and $x < l_f/2$, respectively. For the cases with large values of $l_f/l_0$, $F_1(x) = -F_2(x) \approx 0$ at $x = l_f$, and $F_2(x) = -F_3(x) \approx 0$ at $x = 0$. Reminding the force equilibrium illustrated in Fig. 2, we can naturally deduce that the shear strains $\gamma_{13}$ and $\gamma_{12}$ are also eventually converged to the values of $\gamma_{23}$ and $\gamma_{13}$ at regions of $x > l_f/2$ and $x < l_f/2$, respectively. Fig. 4b-d confirm these relationships between the shear strain distributions in different layers. Noted that the maximum tensile force and interlayer shear strain are all located at the edges of folded layers that are not connected with adjacent layers, implying where the tension-induced structure failure would initiate. The clarifications of detailed load transfer behaviors in each folded layer establish a solid foundation for the studies on the overall elastic properties of SF-2DNMs in the following sections.

*3.2 Effective Young's modulus*

The effective Young's modulus for the whole structure can be defined as the ratio between the effective tensile stress $F_0/3h$, where $3h$ is the thickness of RVE, and the effective strain $[u_3(l_f) - u_1(0)]/l_f$, where $[u_3(l_f) - u_1(0)]$ is the total elongation of the RVE. Combining the results in Eq. (4), the effective Young's modulus $E_{\text{eff}}$ is derived as

$$E_{\text{eff}} = \frac{F_0/3h}{[u_3(l_f) - u_1(0)]/l_f} = \frac{D}{h} \cdot \frac{1}{1 + \frac{4n^2 - 4}{n^2 - n + 1} \cdot \left(\frac{l_f}{l_0}\right)^{-1}} \qquad (7)$$

The evolution of $E_{\text{eff}}$ with $l_f/l_0$ is then plotted in Fig. 5a, which is normalized by $D/h$. It should be mentioned that there are some extra restrictions on the variation range of fold length in practical SF-2DNMs. Some previous studies by MD simulation found that the folded structure of SF-2DNMs keeps its thermodynamic stability only when the fold length is larger than a critical value [16, 26]. Therefore, we set the lower bound of fold length at $l_f/l_0 = 2$ based on the critical values reported in literatures. Fig. 5a shows that the $E_{\text{eff}}$ firstly increases with $l_f/l_0$ then gradually converges to a plateau for large $l_f/l_0$ and finally gets saturated at $l_f/l_0 \approx 100$. The ultimate stable value of $E_{\text{eff}}$ can be calculated from Eq. (7) as $D/h$, which is independent on the interlayer shear properties of 2DNMs. It indicates that the stiffness of SF-2DNMs with very large fold length is solely contributed by the in-plane tensile properties of layers of 2DNMs.

This evolution can be interpreted by the relatively suppressed load transfer between adjacent layers with the increasing $l_f/l_0$, as shown in our foregoing discussions.

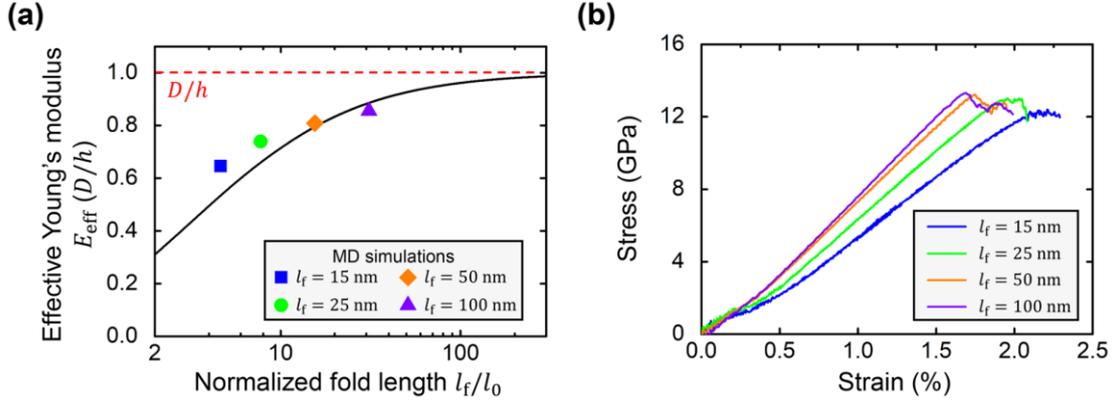

**Fig. 5 Predictions of effective Young's modulus of self-folded two-dimensional nanomaterials (SF-2DNMs) with validations by molecular dynamics (MD) simulations.** (**a**) Evolution of effective Young's modulus with the normalized fold length $l_f/l_0$. The black solid line represents the description made by our theoretical model. MD simulation results of self-folded graphene are also indicated as the colored dots for comparisons. (**b**) The stress-strain curves obtained from MD simulations for self-folded graphene under tension.

To verify our theoretical predictions of elastic properties, we also use the MD simulations to calculate the effective Young's modulus for the SF-2DNM models with $l_f =$ 15, 25, 50 and 100 nm. Graphene is adopted as a typical example of 2DNMs in our simulations. The simulation cell is constructed with the same tri-folded configuration of A-A stacked graphene layers as that in the RVE model. Modified AIREBO potential proposed by Shen et al. [27] is employed in our simulations for better descriptions of the interactions between graphene layers. Periodic boundary conditions are set for all three directions of the cell. The tension is implemented by increasing the length of cell and remapping the positions of all the atoms accordingly. The temperature of system is controlled at 1 K to avoid unnecessary thermal fluctuations. Then the $E_{\text{eff}}$ can be fitted from the linear regions in the stress-strain curves obtained from MD simulations as shown in Fig. 5b. Based on the mechanical properties of single-layer graphene revealed by previous MD studies [27] using the same potential, such as $E = 1025$ GPa, $G = 3.8$ GPa, $h = 0.34$ nm, and assuming $t = h = 0.34$ nm [4, 28], the simulation results of $E_{\text{eff}}$ can be normalized into the same forms of our analytical results, which are plotted in Fig. 5a as well. The theoretical predictions agree well with these effective Young's modulus of SF-graphene calculated from MD simulations. We note that our model produces relatively larger deviations on $E_{\text{eff}}$ for SF-2DNMs with smaller fold length. That may be attributed to the overlook of the connections between adjacent layers in our mechanical modeling. The fold length of SF-2DNMs is not large enough to completely omit the role of these secondary structures in the overall elastic properties.

*3.3 Failure mode and strength under tension*

From the above discussions on the distributions of in-plane tensile force and interlayer shear strain, it has been known that the failure of overall folded structure nucleates in the edges of folded layers without connections with adjacent layers. The SF-2DNM could fail in two modes, one is the fracture of 2DNM layer when the in-plane stress exceeds the intrinsic strength of 2DNM $\sigma_{\text{cr}}$ (called mode G), the other is the failure of interlayer interaction when the critical shear $\gamma_{\text{cr}}$ is achieved between layers (called mode I). For the design purpose of both excellent strength and ductility, the SF-2DNMs are preferred to fall in mode I instead of mode G in practical applications [16]. The failure modes for different SF-2DNMs are determined by not only the intralayer and interlayer material properties but also the geometrical characteristics of folded structure. The failure criterial of model G and mode I can be written as

$$\begin{cases} \dfrac{F_0 l_0}{Dh} \cdot \dfrac{n^2-1}{n^2-n+1} > \gamma_{\text{cr}}, & \text{mode I} \\ \dfrac{F_0}{t} > \sigma_{\text{cr}}, & \text{mode G} \end{cases} \tag{8}$$

For the SF-2DNM, the tensile strengths $\sigma_s$ is calculated as $F_0/3h$. The overall structure should follow the fracture mode with lower values of $\sigma_s$ based on each criterion. Introducing two nondimensional controlling parameters $k_1 = l_f/l_0$ and $k_2 = \sigma_{cr} l_0 t / \gamma_{cr} Dh$, the tensile strength can be expressed as

$$\sigma_s = \begin{cases} \dfrac{\gamma_{\text{cr}} D}{l_0} \cdot \dfrac{n^2-n+1}{3n^2-3}, & k_2 \geq \dfrac{e^{2k_1}-1}{e^{2k_1}-e^{k_1}+1} \text{ (mode I)} \\ \dfrac{\sigma_{\text{cr}} t}{3h}, & k_2 < \dfrac{e^{2k_1}-1}{e^{2k_1}-e^{k_1}+1} \text{ (mode G)} \end{cases} \tag{9}$$

Then we can illustrate a diagram of two failure modes with two axes of $k_1$ and $k_2$, as shown in Fig. 6a. In the diagram, the boundary between the regions of two modes rapidly converges to a horizontal line $k_2 = 1$, indicating that the failure mode of SF-2DNM is only determined by the value of $k_2$ when fold length is relatively large. For this case, if $k_2 > 1$ satisfies for a certain 2DNM, its self-folded assembly can be gradually unfolded by the progressive interlayer failure under tension, which is exactly required for the achievement of ultrahigh ductility of 2DNMs. In contrast, for $k_2 < 1$, the material layers of SF-2DNMs would be firstly fractured under tension, and result in the drastic deterioration of bearing capacity and relatively low ductility as well as toughness, which should be avoided in the material design by the folding strategy. We can substitute the material properties and structural parameters of an arbitrary SF-2DNM into this diagram to estimate its failure mode. Our model would provide helpful guidelines for the material selection and structure adjustment for SF-2DNMs.

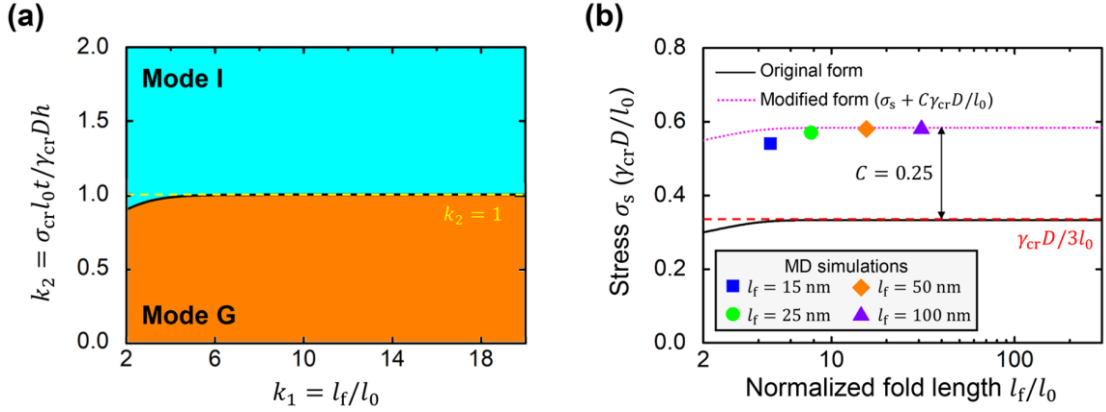

**Fig. 6 Predictions of failure modes and tensile strength of self-folded two-dimensional nanomaterials (SF-2DNMs) with validations by molecular dynamics (MD) simulations.** (**a**) The diagram of different failure modes for SF-2DNMs. Mode I and mode G indicate the failure of interlayer interactions and fracture of 2DNM layers, respectively. $k_1$ and $k_2$ are two nondimensional parameters controlling the failure mode of SF-2DNMs. (**b**) Evolution of tensile strength with the normalized fold length $l_f/l_0$. The black solid line represents the description made by the original form of our theoretical model. MD simulation results of self-folded graphene are also indicated as the colored dots for comparisons. The pink dotted line shows the predictions by the modified model that fits the enhancement on tensile strength for self-folded graphene. Here the constant $C$ is fitted as 0.25.

Moreover, MD simulations are performed for validations in this section. Taking graphene as an example as well. The value of $k_2$ for the SF-graphene can be calculated as ~5 using $\gamma_{cr} = 0.21$ [29], $\sigma_{cr} = 100$ GPa [28] and other parameters listed above. According to the diagram, the SF-graphene would definitely fail as the mode I under tension, which is confirmed by our MD simulations. Under mode I, the prediction of tensile strength $\sigma_s$ using the Eq. (9) is plotted in Fig. 6b with the normalization by

$\gamma_{\mathrm{cr}} D/l_0$. It is observed that $\sigma_\mathrm{s}$ quickly approaches the convergence of $\gamma_{\mathrm{cr}} D/3l_0$ with the increasing fold length $l_\mathrm{f}/l_0$. Then the tensile strengths of SF-graphene obtained from MD simulations are indicated in Fig. 6b following the same normalization. Comparing with theoretical predictions derived from our model, these simulated results of $\sigma_\mathrm{s}$ exhibit similar saturation trends but higher quantities with the growth of $l_\mathrm{f}/l_0$. The underestimate of tensile strength has been also reported and well demonstrated by He et al. [30] in the mechanical model developed for layer-by-layer stacked assembly of graphene, which is attributed to the edge effect on the interlayer shear between graphene layers [31, 32]. The edge effect only makes a slight influence on the Young's modulus but apparently enhances the tensile strength, causing the good coincidence on $E_{\mathrm{eff}}$ but significant difference on $\sigma_\mathrm{s}$ as shown in Fig. 5a and 6b, respectively. He et al. also found that the edge effect would make a constant contribution to the total strength when the length of graphene layer is relatively large [30]. That is consistent with the observation of uniform quantitative deviations between theoretical and simulated results for the four cases in Fig. 6b with different fold lengths. Therefore, it is inspired that we can modify our model by adding a constant $C$ into the normalized expression of $\sigma_\mathrm{s}$ for the SF-graphene. The modified curve of $\sigma_\mathrm{s}$ with $C$ fitted as 0.25 are plotted in Fig. 6b, which shows good agreement with those results obtained from MD simulations. Based on the modified expression of $\sigma_\mathrm{s}$ for mode I, the diagram of the failure mode should be reproduced correspondingly as well. He et al. further reported that the edge effect is not observed for all 2DNMs, for example, the stacked assemblies of $MoS_2$ [8]. Thus the modification of $\sigma_\mathrm{s}$ is not necessary for these SF-2DNMs, which elastic properties are believed to be well described by the original forms of our model.

## 4. Conclusions

In summary, we develop a theoretical model for the elastic properties of SF-2DNMs based on the shear-lag analysis with the validations of MD simulations. This model demonstrates the effects of structural geometries on the overall elastic properties of SF-2DNMs. The load transfer behaviors in SF-2DNMs are investigated with our model. It is found that the interlayer shearing gradually concentrates on limited regions near the edges of folded layers as the normalized fold length $l_\mathrm{f}/l_0$ increases. The Young's modulus and tensile strength are also predicted by this model, which firstly increase with the $l_\mathrm{f}/l_0$ then finally converge to constants. The diagram of different failure modes is obtained from this model as well. Moreover, this model can be slightly modified to consider the enhancement effect on the tensile strength for certain 2DNMs, such as graphene. The structure-property relationships revealed by this mechanical model would provide effective guidelines for the utilization of the folding strategy in the design of 2DNM-based materials with both excellent strength and ductility.

**Declaration of Competing Interest**

The authors declare that they have no known competing financial interests or personal relationships that could have appeared to influence the work reported in this paper.

**Acknowledgment**

This work was supported by the National Natural Science Foundation of China (Grant No. 11972226).

**References**

[1] A.K. Geim, K.S. Novoselov, The rise of graphene, Nature Materials 6 (2007) 183-191.

[2] H. Zhang, M. Chhowalla, Z. Liu, 2D nanomaterials: graphene and transition metal dichalcogenides, Chemical Society Reviews 47 (2018) 3015-3017.

[3] H. Zhang, Ultrathin Two-Dimensional Nanomaterials, ACS Nano 9 (2015) 9451-9469.

[4] C. Lee, X. Wei, J.W. Kysar, J. Hone, Measurement of the elastic properties and intrinsic strength of monolayer graphene, science 321 (2008) 385-388.

[5] X. Li, T. Yang, Y. Yang, J. Zhu, L. Li, F.E. Alam, X. Li, K. Wang, H. Cheng, C.-T. Lin, Y. Fang, H. Zhu, Large-Area Ultrathin Graphene Films by Single-Step Marangoni Self-Assembly for Highly Sensitive Strain Sensing Application, Advanced Functional Materials 26 (2016) 1322-1329.

[6] X. Yu, M.S. Prévot, N. Guijarro, K. Sivula, Self-assembled 2D WSe2 thin films for photoelectrochemical hydrogen production, Nature Communications 6 (2015) 7596.

[7] H. Chen, M.B. Müller, K.J. Gilmore, G.G. Wallace, D. Li, Mechanically Strong, Electrically Conductive, and Biocompatible Graphene Paper, Advanced Materials 20 (2008) 3557-3561.

[8] D.A. Dikin, S. Stankovich, E.J. Zimney, R.D. Piner, G.H.B. Dommett, G. Evmenenko, S.T. Nguyen, R.S. Ruoff, Preparation and characterization of graphene oxide paper, Nature 448 (2007) 457-460.

[9] Z. Xu, C. Gao, Graphene fiber: a new trend in carbon fibers, Materials Today 18 (2015) 480-492.

[10] Y. Niu, R. Wang, W. Jiao, G. Ding, L. Hao, F. Yang, X. He, MoS2 graphene fiber based gas sensing devices, Carbon 95 (2015) 34-41.

[11] M. Yang, N. Zhao, Y. Cui, W. Gao, Q. Zhao, C. Gao, H. Bai, T. Xie, Biomimetic Architectured Graphene Aerogel with Exceptional Strength and Resilience, ACS Nano 11 (2017) 6817-6824.

[12] Z. Tan, M. Zhang, C. Li, S. Yu, G. Shi, A General Route to Robust Nacre-Like Graphene Oxide

Films, ACS Applied Materials & Interfaces 7 (2015) 15010-15016.

[13] L. Peng, Z. Xu, Z. Liu, Y. Guo, P. Li, C. Gao, Ultrahigh Thermal Conductive yet Superflexible Graphene Films, Advanced Materials 29 (2017) 1700589.

[14] S. Wang, Y. Gao, A. Wei, P. Xiao, Y. Liang, W. Lu, C. Chen, C. Zhang, G. Yang, H. Yao, T. Chen, Asymmetric elastoplasticity of stacked graphene assembly actualizes programmable untethered soft robotics, Nature Communications 11 (2020) 4359.

[15] Y. Xiao, Z. Xu, Y. Liu, L. Peng, J. Xi, B. Fang, F. Guo, P. Li, C. Gao, Sheet Collapsing Approach for Rubber-like Graphene Papers, ACS Nano 11 (2017) 8092-8102.

[16] X. Jia, Z. Liu, E. Gao, Bio-inspired self-folding strategy to break the trade-off between strength and ductility in carbon-nanoarchitected materials, npj Computational Materials 6 (2020) 13.

[17] A. Nova, S. Keten, N.M. Pugno, A. Redaelli, M.J. Buehler, Molecular and Nanostructural Mechanisms of Deformation, Strength and Toughness of Spider Silk Fibrils, Nano Letters 10 (2010) 2626-2634.

[18] K. Kim, Z. Lee, B.D. Malone, K.T. Chan, B. Alemán, W. Regan, W. Gannett, M.F. Crommie, M.L. Cohen, A. Zettl, Multiply folded graphene, Physical Review B 83 (2011) 245433.

[19] H.L. Cox, The elasticity and strength of paper and other fibrous materials, British Journal of Applied Physics 3 (1952) 72-79.

[20] B. Ji, H. Gao, Mechanical properties of nanostructure of biological materials, Journal of the Mechanics and Physics of Solids 52 (2004) 1963-1990.

[21] Y. Liu, B. Xie, Z. Zhang, Q. Zheng, Z. Xu, Mechanical properties of graphene papers, Journal of the Mechanics and Physics of Solids 60 (2012) 591-605.

[22] E. Gao, Y. Cao, Y. Liu, Z. Xu, Optimizing Interfacial Cross-Linking in Graphene-Derived Materials, Which Balances Intralayer and Interlayer Load Transfer, ACS Applied Materials & Interfaces 9 (2017) 24830-24839.

[23] K. Wu, Z. Song, L. He, Y. Ni, Analysis of optimal crosslink density and platelet size insensitivity in graphene-based artificial nacres, Nanoscale 10 (2018) 556-565.

[24] Z. He, Y. Zhu, J. Xia, H. Wu, Optimization design on simultaneously strengthening and toughening graphene-based nacre-like materials through noncovalent interaction, Journal of the Mechanics and Physics of Solids 133 (2019) 103706.

[25] X. Wei, T. Filleter, H.D. Espinosa, Statistical shear lag model – Unraveling the size effect in

hierarchical composites, Acta Biomaterialia 18 (2015) 206-212.

[26] A. Wei, H. Ye, Y. Gao, F. Guo, Unveiling the mechanism of structure-dependent thermal transport behavior in self-folded graphene film: a multiscale study, Nanoscale 12 (2020) 24138-24145.

[27] Y. Shen, H. Wu, Interlayer shear effect on multilayer graphene subjected to bending, Applied Physics Letters 100 (2012) 101909.

[28] H. Zhao, K. Min, N.R. Aluru, Size and Chirality Dependent Elastic Properties of Graphene Nanoribbons under Uniaxial Tension, Nano Letters 9 (2009) 3012-3015.

[29] Y. Li, W. Zhang, B. Guo, D. Datta, Interlayer shear of nanomaterials: Graphene—graphene, boron nitride—boron nitride and graphene—boron nitride, Acta Mechanica Solida Sinica 30 (2017) 234-240.

[30] Z. He, Y. Zhu, H. Wu, Edge effect on interlayer shear in multilayer two-dimensional material assemblies, International Journal of Solids and Structures 204-205 (2020) 128-137.

[31] Z. Guo, T. Chang, X. Guo, H. Gao, Thermal-Induced Edge Barriers and Forces in Interlayer Interaction of Concentric Carbon Nanotubes, Physical Review Letters 107 (2011) 105502.

[32] Z. Guo, T. Chang, X. Guo, H. Gao, Mechanics of thermophoretic and thermally induced edge forces in carbon nanotube nanodevices, Journal of the Mechanics and Physics of Solids 60 (2012) 1676-1687.